\documentclass{moriond}

\usepackage{siunitx}
\DeclareSIUnit\overlinen{b}
\usepackage{microtype}

\def\Journal#1#2#3#4{{#1} {\bf #2}, #3 (#4)}

\def\be{\begin{equation}}
\def\ee{\end{equation}}
\def\bea{\begin{eqnarray}}
\def\eea{\end{eqnarray}}

\newcommand{\Photo}{\includegraphics[height=35mm]{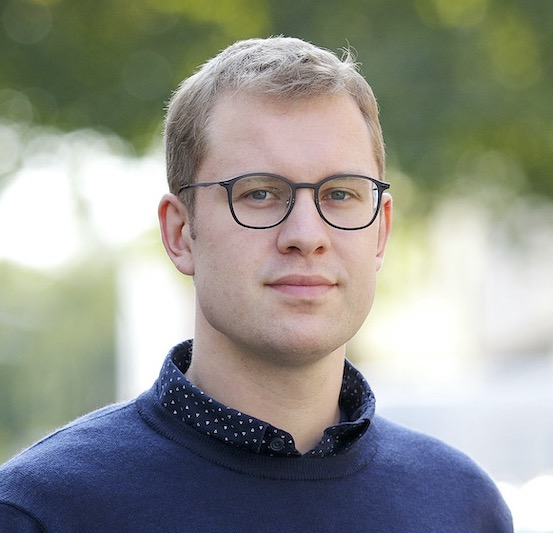}}

\begin{document}
\vspace*{4cm}
\title{Lepton flavour universality tests at LHCb}

\author{Alex Seuthe \\ On behalf of the LHCb collaboration}

\address{TU Dortmund University, Dortmund, Germany}

\maketitle\abstracts{The LHCb experiment at the Large Hadron Collider specialises in high-precision measurements of flavour physics with hadrons containing $b$ and $c$ quarks. Lepton flavour universality tests provide an accurate and clear approach to scrutinising the Standard Model of particle physics. These proceedings report recent lepton flavour universality tests performed by the LHCb collaboration. For $b \to c \ell \nu$ transitions, it includes the first simultaneous measurement of $R_{D^*}$ and $R_{D^{0}}$ at a hadron collider, an updated $R_{D^{*}}$ measurement with hadronic $\tau$ decays using 2015-2016 data, and recent measurements of $R_{J/\psi}$ and $R_{\Lambda_c}$. In $b \to s \ell^+ \ell^-$ transitions, the most recent measurements of $R_{pK}$, $R_{K^{*+}}$, and $R_{K^0_{\mathrm{S}}}$ ratios are presented. Moreover, the simultaneous measurement of $R_K$ and $R_{K^*}$ is reported.}

\section{Introduction}
According to the Standard Model (SM) of particle physics, the coupling of the electroweak gauge bosons to leptons is flavour universal (LFU). Possible new physics (NP) particles could cause a violation of LFU. Various LFU tests have been performed by LHCb studying tree-level $b \to c \ell \nu$ and rare loop-level $b \to s \ell^+ \ell^-$ transitions.

\section{Lepton flavour universality tests with \texorpdfstring{$b \to c \ell \nu$}{b→cℓν} decays}
LFU tests with $b \to c \ell \nu$ decays measure the ratios \mbox{$R_{H_c} = \frac{\mathcal{B}(H_b \to H_c \tau \nu_\tau)}{\mathcal{B}(H_b \to H_c \mu \nu_\mu)}$}, where $H_b$ is a hadron containing a $b$ quark and $H_c$ containing a $c$ quark. The advantages of the ratio approach include the removal of the dependency on CKM matrix element $|V_{cb}|$ and reducing both experimental and theoretical uncertainties. $R_{H_c}$ can be sensitive to NP couplings involving the third lepton generation, with possible new particles. The $\tau$ can further decay through two main channels, with  $\tau^- \to \pi^- \pi^+ \pi^- (\pi^0) \nu_\tau$ (\textit{hadronic}) and with $\tau^- \to \mu^- \overline{\nu}_\mu \nu_\tau$ (\textit{muonic}).

\subsection{Simultaneous measurement of \texorpdfstring{$R_{D^*}$ and $R_{D^{0}}$}{R(D*) and R(D⁰)}}
In the first simultaneous measurement of $R_{D^*}$ and $R_{D^0}$ using LHCb Run 1 data~\cite{ref03}, the ratios $R_{D^{(*)}} = \frac{\mathcal{B}(\overline{B} \to D^{(*)} \tau^- \overline{\nu}_\tau)}{\mathcal{B}(\overline{B} \to D^{(*)} \mu^- \overline{\nu}_\mu)}$ are measured, with $\tau^- \to \mu^- \overline{\nu}_\mu \nu_\tau$. Earlier, $R_{D^*}$ was measured with Run 1 $D^{*+}\mu^-$ data, showing a $2.1\sigma$ deviation from the SM prediction~\cite{ref04}. The recent analysis benefits from adding a $D^{0} \mu^-$ sample. Reconstruction challenges include three neutrinos in the final state, leading to a broad invariant mass distribution. A template fit is performed using variables $m^2_\text{miss.} = (p_B - p_{D^{(*)}} - p_\mu)^2$, $q^2 =(p_B - p_{D^{(*)}})^2$, and $E_\mu$. The analysis yields $R_{D^*} = 0.281 \pm 0.018 \text{(stat.)} \pm 0.024 \text{(syst.)}$ and $R_{D^0} = 0.441 \pm 0.060 \text{(stat.)} \pm 0.066 \text{(syst.)}$, with a correlation of $\rho = -0.43$. Limited data and simulation samples are the main sources of systematic uncertainty. These results indicate a $1.9 \sigma$ agreement with the SM. The new preliminary average reveals a slight decrease in $R_{D^*}$ and a slight increase in $R_{D^0}$ \cite{ref12}.

\subsection{Measurement of \texorpdfstring{$R_{D^{*}}$ with hadronic $\tau$ decays}{R(D*) with hadronic τ decays}}
\begin{figure}
	\centering
	\includegraphics[width=0.6\textwidth]{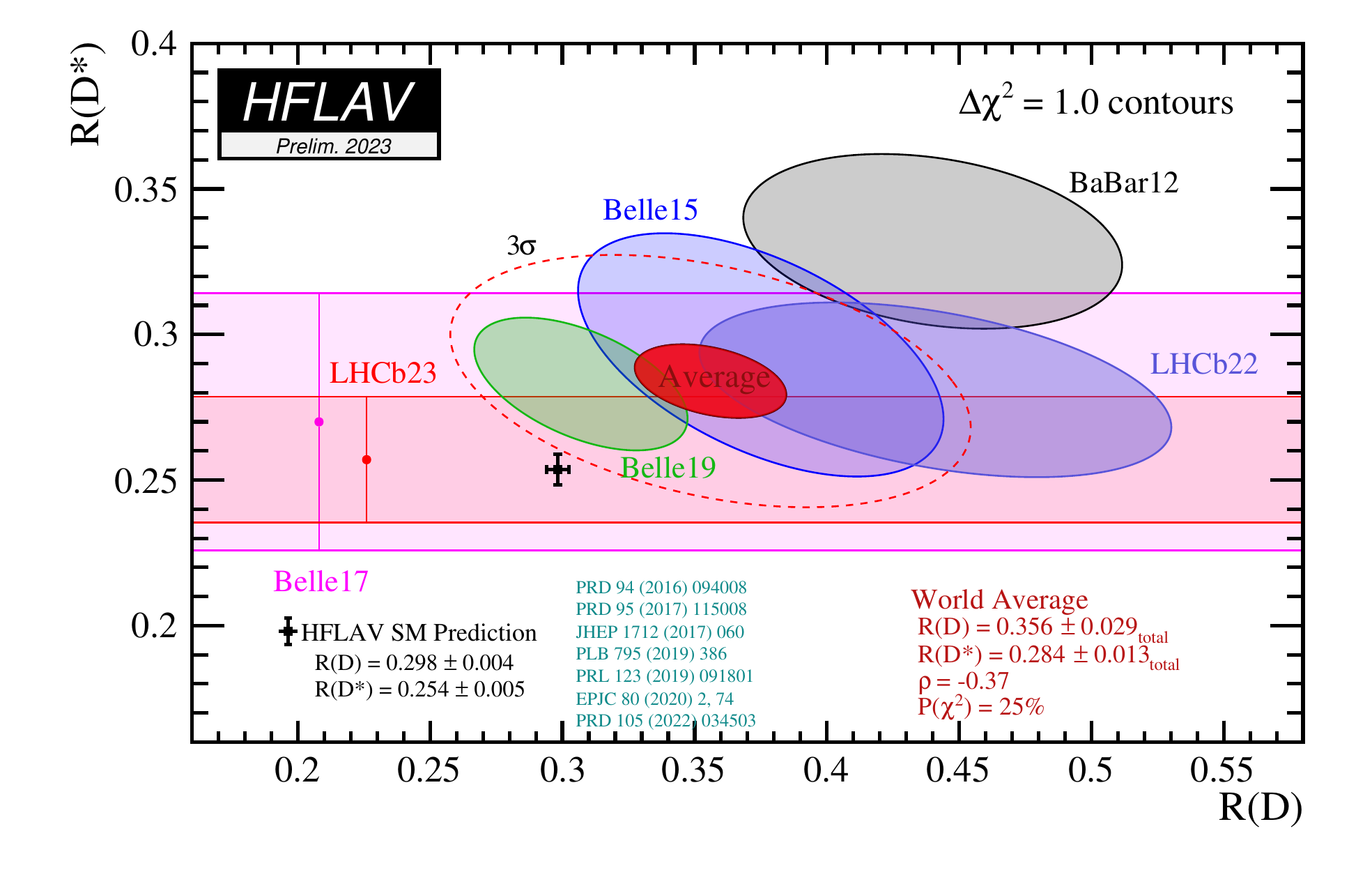}
	\caption{HFLAV preliminary average of $R_{D^0}$ and $R_{D^*}$ for winter 2023\protect\cite{ref12}.}
	\label{fig:HLFAVaverage}
\end{figure}

The recent measurement of $R_{D^*}$ with hadronic $\tau$ decays~\cite{ref05} adds the 2015-2016 data to the LHCb Run 1 analysis~\cite{ref06,ref07}. For the determination of $R_{D^*}$, the decay $B^0 \to D^{-} 3 \pi^{\pm}$ with the same visible three-prong final state is chosen as normalisation mode. The ratio is determined as \mbox{$R_{D^*} = \mathcal{K}_{D^*} \cdot \frac{\mathcal{B}(B^0 \to D^{-} 3 \pi^{\pm})}{\mathcal{B}(B^0 \to D^{-} \mu^+ \nu_\mu)}$} with \mbox{$\mathcal{K}_{D^*} = \frac{\mathcal{B}(B^0 \to D^{-} \tau^+ \nu_\tau)}{\mathcal{B}(B^0 \to D^{-} 3 \pi^{\pm})} = \frac{N_\text{Sig.}}{N_\text{Norm.}} \cdot \frac{\epsilon_\text{Norm.}}{\epsilon_\text{Sig.}} \cdot \frac{1}{\mathcal{B}(\tau^+ \to 3 \pi^{\pm} (\pi^0)\overline{\nu}_\tau)}$}. The branching fractions of the decays $B^0 \to D^{-} 3 \pi^{\pm}$ and $B^0 \to D^{-} \mu^+ \nu_{\mu}$ are taken from external inputs.  With a 3D template fit the result of \mbox{$R_{D^*} = 0.257 \pm 0.012 \text{(stat.)} \pm 0.014 \text{(syst.)} \pm 0.012 \text{(ext.)}$} is extracted. Fig.~\ref{fig:HLFAVaverage} shows the new preliminary world average values are \mbox{$R_{D^*} = 0.284 \pm 0.013$} and \mbox{$R_{D^0} =  0.356 \pm 0.029$}. The global discrepancy with the SM is at $3.2 \sigma$.

\subsection{Measurements of \texorpdfstring{$R_{J/\psi}$ and $R_{\Lambda_c}$}{R(J/ψ) and R(Λc)} with LHCb Run 1 data}
Measurements of the ratios $R_{J/\psi}$ and $R_{\Lambda_c}$ represent the first LFU tests with $B_c^+$ mesons and baryonic $b \to c \ell \nu$ decays, respectively~\cite{ref08,ref10}. The $R_{J/\psi}$ ratio is defined as $R_{J/\psi} = \frac{\mathcal{B}(B_c^+ \to J/\psi \tau^+ \nu_{\tau})}{\mathcal{B}(B_c^+ \to J/\psi \mu^+ \nu_\mu)}$, with $\tau^+ \to \mu^+ \nu_\mu \overline{\nu}_{\tau}$. Form factors are determined from fits to data, and the main systematic uncertainties of the final result arise from the limited sample sizes and the form factors themselves. The result $R_{J/\psi} = 0.71 \pm 0.17(\text{stat.}) \pm 0.18(\text{syst.})$ lies $2 \sigma$ above the SM prediction of $0.2583 \pm 0.0038$~\cite{ref09}.
$R_{\Lambda_c}$ is defined as $R_{\Lambda_c} = \frac{\mathcal{B}(\Lambda_b \to \Lambda_c^+ \tau^- \overline{\nu}_\tau)}{\mathcal{B}(\Lambda_b \to \Lambda_c^+ \mu^- \overline{\nu}_\mu)}$, with $\tau^- \to \pi^- \pi^+ (\pi^0) \nu_\tau$. The largest systematic uncertainty stems from background template shapes. The result of \mbox{$R_{\Lambda_c} = 0.242 \pm 0.026(\text{stat.}) \pm 0.040(\text{syst.}) \pm 0.059(\text{ext.})$} is in $1 \sigma$ agreement with the SM prediction of $0.324 \pm 0.004$~\cite{ref11}.

\section{Lepton flavour universality tests with \texorpdfstring{\mbox{$b \to s \ell^+ \ell^-$}}{b→sℓ⁺ℓ⁻} decays}
Rare $b \to s \ell^+ \ell^-$ decays occur only at the loop level and exhibit sensitivity to NP. Branching fraction ratios of decays with various hadronic final states are employed to measure the ratios $R_H = \frac{\mathcal{B}(B\to H \mu^+ \mu^-)}{\mathcal{B}(B\to H e^+ e^-)}$ in different regions of $q^2 = m^{2}(\ell^+ \ell^-)$. Except for different masses of the leptons involved and kinematic effects, these ratios are precisely expected to be unity~\cite{ref13}. The measurements are performed as double ratios with the $J/\psi$ charmonium mode to cancel out systematic uncertainties.

\subsection{Simultaneous measurement of \texorpdfstring{$R_K$ and $R_{K^{*}}$}{R(K) and R(K*)}}
\begin{figure}
	\centering
	\includegraphics[width=0.6\textwidth]{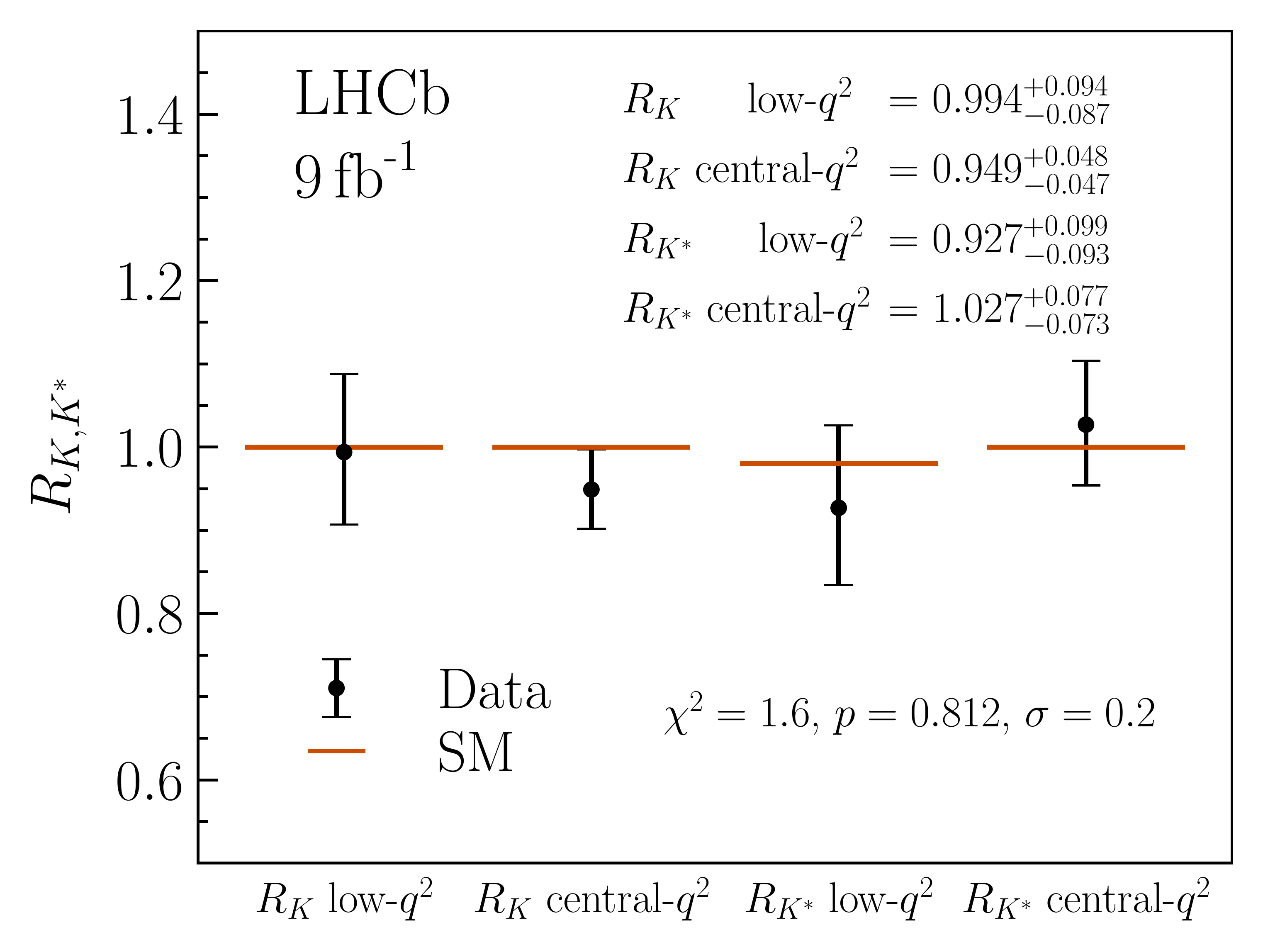}
	\caption{Summary of results for the measurement of $R_K$ and $R_{K^*}$\protect\cite{ref14}.}
	\label{fig:RXsummary}
\end{figure}

The first simultaneous measurement of $R_K$ and $R_{K^*}$ is performed with the full Run 1 and Run 2 LHCb data sets in two $q^2$ regions: \textit{low} ($q^2 \in [0.1,1.1] \, \si{\giga \electronvolt^2 / c^4}$) and \textit{central} ($q^2 \in [1.1,6.0] \, \si{\giga \electronvolt^2 / c^4}$)~\cite{ref14,ref15}. Cross-feed background between the two decay modes is constrained in fits to data, and the simulation is calibrated with $B^{+/0} \to K^{+/*0} J/\psi (\to \ell^+ \ell^-)$ decays, decoupling it from the normalisation mode and enabling cross-validation. To reduce mis-ID backgrounds, stringent particle identification requirements are applied for both leptons and hadrons. Multivariate classifiers are employed against partially reconstructed and combinatorial backgrounds, while physical backgrounds are vetoed. Efficiency-calibration cross-checks are performed using ratios of branching fractions of the $J/\psi$ and $\psi(2S)$ charmonium modes, which are compatible with unity, as expected. When interpreted as a null test, the four measurements are compatible with the SM at $0.2 \sigma$. These statistically dominated results represent the highest precision LFU test with $b \to s \ell^+ \ell^-$ decays to date and supersede previous LHCb measurements. Fig.~\ref{fig:RXsummary} presents a summary of the results.

\subsection{Measurement of \texorpdfstring{$R_{pK}$}{R(pK)}}
LFU is tested with baryons for the first time with the measurement of \mbox{$R_{pK} = \frac{\mathcal{B}(\Lambda_b^0 \to p K^- \mu^+\mu^-)}{\mathcal{B}(\Lambda_b^0 \to p K^- e^+e^-)}$~\cite{ref16}}. It provides sensitivity to the spin dependence of potential NP. Utilising LHCb Run 1 and 2016 data, the measurement yields $R_{pK}^{[0.1,6.0]} = 0.86^{+0.14}_{-0.11} \text{(stat.)} \pm 0.05 \text{(syst.)}$, which agrees with unity within $1 \sigma$.

\subsection{Measurement of \texorpdfstring{$R_{K^{*+}}$ and $R_{K^0_\mathrm{S}}$}{R(K*⁺) and R(K⁰S)}}
The ratios $R_{K^{*+}}$ and $R_{K_\mathrm{S}^0}$ are measured~\cite{ref17} using the Run 1 and 2 LHCb data sets with \mbox{$B^+ \to K^{*+}\ell^+\ell^-$} and \mbox{$B^0 \to K_\mathrm{S}   ^0 \ell^+ \ell^-$} ($\ell = \mu, e$) decays, respectively. With this measurement, the first observation of the $B^+ \to K^{*+} e^+e^-$ and $B^0 \to K_\mathrm{S}^0 e^+e^-$ decays is reported. The results of $R_{K^{*+}}^{[0.045,6.0]} = 0.70^{+0.18}_{-0.13} \text{(stat.)} ^{+0.03}_{-0.04} \text{(syst.)}$ and $R_{K_{\mathrm{S}}^{0}}^{[1.0,6.0]} = 0.66^{+0.20}_{-0.14} \text{(stat.)} ^{+0.02}_{-0.04} \text{(syst.)}$ are found to be compatible with the SM within $1.4 \sigma$ and $1.5 \sigma$, respectively.

\section{Conclusion and prospects}
These proceedings present LFU tests with $b \to c\ell\nu$ transitions, with the first simultaneous measurement of $R_{D^0}$ and $R_{D^*}$ using muonic $\tau$ decays at LHCb and an update of the hadronic measurement of $R_{D^*}$. The global picture for the $R_{D^0}$-$R_{D^*}$ combination remains unchanged with SM tension at the $3\sigma$ level. In the $b \to s \ell^+\ell^-$ transition sector, the first simultaneous measurement of $R_K$ and $R_{K^*}$ currently provides the most precise and accurate LFU test. These results are compatible with the SM within $0.2 \sigma$. However, anomalies in differential branching fractions and angular analyses of the muon modes persist. The LHCb Upgrade I detector is currently being commissioned, with plans to increase the instantaneous luminosity by a factor of 5 and collect approximately $\SI{50}{\femto\overlinen^{-1}}$ of data during Run 3.

\section*{Acknowledgments}
A.S. acknowledges support from the European Research Council (ERC) under the European Union's Horizon 2020 research and innovation programme under grant no. 714536: PRECISION, the German Federal Ministry of Education and Research (BMBF, grant no. 05H21PECL1) within ErUM-FSP T04, and the German Academic Scholarship Foundation (Studienstiftung des deutschen Volkes).

\end{document}